\documentclass[10pt,fleqn,twoside]{article}
\usepackage{espcrc2}
\usepackage{graphicx,feynarts}
\usepackage[figuresright]{rotating}

\input{paperdef}

\begin{document}

\thispagestyle{empty}
\setcounter{page}{0}
\def\thefootnote{\fnsymbol{footnote}}

\begin{flushright}
\mbox{}
\end{flushright}

\vspace{1cm}

\begin{center}

{\large\sc {\bf Electroweak Precision Physics\\[.5em] 
from Low to High Energies}}
\footnote{plenary talk given at the {\em Lepton Photon 07}, 
August 2007, Daegu, Korea}

\vspace{1cm}

{\sc 

S.~Heinemeyer%
\footnote{
email: Sven.Heinemeyer@cern.ch}
}

\vspace*{1cm}

{\it
Instituto de Fisica de Cantabria (CSIC-UC), 
Santander,  Spain
}
\end{center}

\vspace*{0.2cm}

\BC {\bf Abstract} \EC
Electroweak precision observables (EWPO) can give valuable information
about the last unknown paramter of the Standard Model (SM), the
Higgs-boson mass $\MHSM$. EWPO can also restrict the parameter space of
new physics models (NPM) such as the Minimal Supersymmetric Standard
Model (MSSM). We review the respective constraints from the $W$~boson
mass, the effective leptonic mixing angle, the anomalous magnetic moment
of the muon and electric dipole moments. Within the MSSM also the
lightest Higgs-boson mass, $\Mh$, is discussed as a precision observable. The
EWPO, supplemented with $B$~physics observables and astrophysical data
can be used to determine indirectly the preferred mass scales of
Supersymmetry and~$\Mh$.

\def\thefootnote{\arabic{footnote}}
\setcounter{footnote}{0}

\newpage


\hyphenation{re-commend-ed Post-Script}

\graphicspath{{figs/}}

\title{Electroweak Precision Physics from Low to High Energies}

\author
    {
      S.~Heinemeyer\address[Santander]
      {
        Instituto de Fisica de Cantabria (CSIC-UC), Santander, Spain
      }
    }

\begin{abstract}
Electroweak precision observables (EWPO) can give valuable information
about the last unknown paramter of the Standard Model (SM), the
Higgs-boson mass $\MHSM$. EWPO can also restrict the parameter space of
new physics models (NPM) such as the Minimal Supersymmetric Standard
Model (MSSM). We review the respective constraints from the $W$~boson
mass, the effective leptonic mixing angle, the anomalous magnetic moment
of the muon and electric dipole moments. Within the MSSM also the
lightest Higgs-boson mass, $\Mh$, is discussed as a precision observable. The
EWPO, supplemented with $B$~physics observables and astrophysical data
can be used to determine indirectly the preferred mass scales of
Supersymmetry and~$\Mh$. 

\end{abstract}

\maketitle


\section{Introduction}

The Standard Model (SM)~\cite{sm} cannot be the ultimate theory of
particle physics. While describing direct experimental data reasonably
well, it fails to include gravity, it does not provide cold dark matter,
and it has no solution to the hierarchy problem, i.e.\ it does not have
an explanantion for a Higgs-boson mass at the electroweak scale. 
On wider grounds, the SM does not have an explanation for the three
generations of fermions or their huge mass hierarchies.
In order to overcome (at least some of) the above problems, many new
physics models (NPM) have been proposed in the last
decades~\cite{mssm,thdm,lhm,edm}. 

Theories based on Supersymmetry (SUSY)~\cite{mssm} are widely
considered as the theoretically most appealing extension of the
SM. They are consistent with the approximate
unification of the gauge coupling constants at the GUT scale and
provide a way to cancel the quadratic divergences in the Higgs sector
hence stabilizing the huge hierarchy between the GUT and the Fermi
scales. Furthermore, in SUSY theories the breaking of the electroweak
symmetry is naturally induced at the Fermi scale, and the lightest
supersymmetric particle can be neutral, weakly interacting and
absolutely stable, providing therefore a natural solution for the dark
matter problem.
SUSY predicts the existence of scalar partners $\tilde{f}_L,
\tilde{f}_R$ to each SM chiral fermion, and spin--1/2 partners to the
gauge bosons and to the scalar Higgs bosons. 
The Higgs sector of the Minimal Supersymmetric Standard Model (MSSM) 
with two scalar doublets accommodates five physical Higgs bosons. In
lowest order these are the light and heavy $\cp$-even $h$
and $H$, the $\cp$-odd $A$, and the charged Higgs bosons $H^\pm$.
Higher-order contributions yield large corrections to the masses
and couplings. They can also induce $\cp$-violation leading to
mixing between $h,H$ and $A$ in the case of general complex SUSY breaking
parameters.  

Other (non-SUSY) NPM comprise Two Higgs Doublet Models
(THDM)~\cite{thdm}, little Higgs models~\cite{lhm}, or models with
(large, warped, \ldots) extra dimensions~\cite{edm}. 
In specific examples given later, we will mostly focus on the MSSM.
However, the MSSM should be seen as a representative for a NPM. The
reader may insert her/his favorite model.

So far, the direct search for NPM particles has not been successful.
One can only set lower bounds of ${\cal O}(100)$~GeV on
their masses~\cite{pdg}. The search reach will be extended in various
ways in the ongoing Run~II at the upgraded Fermilab
Tevatron~\cite{TevFuture}. 
The LHC~\cite{atlas,cms} and the $e^+e^-$ International Linear Collider
(ILC)~\cite{teslatdr,orangebook,acfarep} have very good prospects for
exploring NPM at the TeV scale, which is favoured from naturalness
arguments. From the interplay of both machines detailed
information on many NPM can be expected in this case~\cite{lhcilc}.

Besides the direct detection of NPM particles (and Higgs bosons), 
physics beyond the SM can also be probed by precision observables via the
virtual effects of the additional particles.
Observables (such as particle masses, mixing angles, asymmetries etc.)
that can be predicted within a certain model and thus depend sensitively
on the other model parameters constitute a test of the model on the
quantum level. Various models predict different values of the same
observable due to their different particle content and
interactions. This permits to distinguish between e.g.\ the SM and a NPM
via precision observables. However, 
this requires a very high precision
of the experimental results as well as of the theoretical predictions.

The wealth of high-precision measurements carried out at LEP, SLC and
the Tevatron~\cite{LEPEWWG,TEVEWWG} as well as the ``Muon $g-2$ Experiment''
(E821)~\cite{g-2exp} and further low-energy experiments (e.g.\ the
search for electric dipole moments (EDM), see below) provide a
powerful tool for testing the electroweak theory and probing indirect
effects of NPM particles.
The most relevant electroweak precision observables (EWPO) in this
context are the $W$~boson mass, $\MW$, the effective leptonic weak
mixing angle, $\sweff$, and the anomalous magnetic moment of the muon, 
$\amu \equiv (g-2)_\mu/2$. In models in which a Higgs-boson mass can be
predicted, it also constitutes a precision observables, most notably the
mass of the lightest $\cp$-even MSSM Higgs boson, $\Mh$~\cite{PomssmRep}.
While the current exclusion bounds on $\Mh$ already allow to constrain
the MSSM parameter space, 
the prospective accuracy for the measurement of the mass of a light 
Higgs boson at the
LHC of about $200 \mev$~\cite{atlas,cms} or at the ILC of even 
$50 \mev$~\cite{teslatdr,orangebook,acfarep,Snowmass05Higgs}
would promote $\Mh$ to a precision observable.


\section{Example: The \boldmath{$W$} boson mass}
\label{sec:mw}

As a prominent example for the interplay of theory and experiment to
perform a theory test at the quantum level serves the prediction of the
$W$~boson mass, 
$\MW$, in the SM and the MSSM. Progress has been achieved over the last
decade in the experimental measurements as well as in the theory
predictions in the SM and in the MSSM. The current experimental
value~\cite{LEPEWWG,TEVEWWG,MWcdf,gruenewald07} 
\BE
\MW^{\rm exp} = 80.398 \pm 25 \gev
\EE
is based on a combination of the LEP results~\cite{LEP4f,lepewwg} and the
latest CDF measurement~\cite{MWcdf,gruenewald07}. 
The experimental measurement of $\MW$ also required substantial theory
input such as cross section evaluations for LEP~\cite{racoonww,yfsww} or
kinematics of $W$ and $Z$ boson decays~\cite{resbos} or the inclusion of
initial and final state photons~\cite{wgrad} at the Tevatron.

Concerning the theory prediction, the $W$~boson mass can be evaluated from
\BE
\MW^2 \KL 1 - \frac{\MW^2}{\MZ^2}\right) = 
\frac{\pi \al}{\sqrt{2} \GF} \left(1 + \De r\KR ,
\label{eq:delr}
\EE
where $\al$ is the fine structure constant and $\GF$ the Fermi constant.
The radiative corrections are summarized in the quantity 
$\De r$~\cite{sirlin}.
Within the SM the one-loop~\cite{sirlin} and the complete two-loop
result has been obtained for 
$\MW$~\cite{MWSMferm2L,MWSMbos2L,drSMgfals,deltarSMgfals,MWSM}. The latter
consists of the fermionic electroweak
two-loop contributions~\cite{MWSMferm2L}, the purely bosonic two-loop
contributions~\cite{MWSMbos2L} and the QCD corrections of
\order{\al\als}~\cite{drSMgfals,deltarSMgfals}. Higher-order QCD
corrections are known at
\order{\al\als^2}~\cite{drSMgfals2,MWSMQCD3LII}. 
Leading electroweak contributions of order
\order{\gf^2 \als \mt^4} and \order{\gf^3 \mt^6} that enter via the
quantity $\De\rho$~\cite{rho} have been calculated in
\citeres{drSMgf3mh0,drSMgf3,drSMgf3MH}. 
The class of four-loop contributions obtained in 
\citere{deltarhoSM4L} give rise to a numerically negligible effect.
The prediction for $\MW$ within the SM
(or the MSSM) is obtained by evaluating $\De r$ in these models and
solving \refeq{eq:delr} for $\MW$. 

Within the MSSM the most precise available result for $\MW$ has been 
obtained in~\citere{MWpope}. Besides the full SM result, for the MSSM it
includes the full set of one-loop
contributions~\cite{deltarMSSM1lA,deltarMSSM1lB,MWpope}   
as well as the corrections of \order{\al\als}~\cite{dr2lA} and of
\order{\al_{t,b}^2}~\cite{drMSSMal2B,drMSSMal2} to the quantity
$\De\rho$; see~\citere{MWpope} for details.

\begin{figure}[htb!]
\begin{center}
\includegraphics[width=.45\textwidth,height=.35\textwidth] 
                {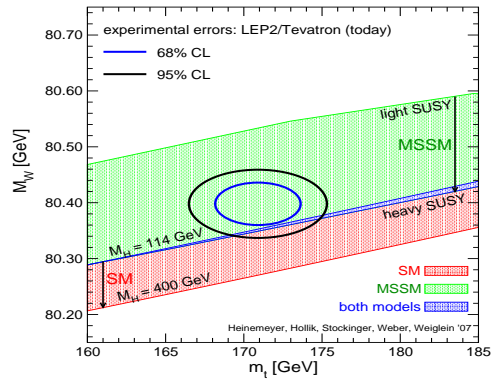}
\vspace{-1em}
\caption{%
Prediction for $\MW$ in the MSSM and the SM (see text) as a function of
$\mt$ in comparison with the present experimental results for $\MW$ and
$\mt$~\cite{MWpope}.
}
\label{fig:MWMTtoday}
\end{center}
\end{figure}

The experimental result and the theory prediction of the SM and the MSSM
are compared in \reffi{fig:MWMTtoday}.%
\footnote{
The plot shown here is an update of 
\citeres{deltarMSSM1lA,PomssmRep,MWpope}.
}%
The predictions within the two models 
give rise to two bands in the $\mt$--$\MW$ plane with only a relatively small
overlap sliver (indicated by a dark-shaded (blue) area in
\reffi{fig:MWMTtoday}).  
The allowed parameter region in the SM (the medium-shaded (red)
and dark-shaded (blue) bands) arises from varying the only free parameter 
of the model, the mass of the SM Higgs boson, from $\MHSM = 114\gev$,
the LEP exclusion bound~\cite{LEPHiggsSM}
(upper edge of the dark-shaded (blue) area), to $400 \gev$ (lower edge
of the medium-shaded (red) area).
The light shaded (green) and the dark-shaded (blue) areas indicate 
allowed regions for the unconstrained MSSM, obtained from scattering the
relevant parameters independently~\cite{MWpope}. 
The decoupling limit with SUSY masses of \order{2 \tev}
yields the lower edge of the dark-shaded (blue) area. Thus, the overlap 
region between
the predictions of the two models corresponds in the SM to the region
where the Higgs boson is light, i.e.\ in the MSSM allowed region 
($\Mh \lsim 135 \gev$~\cite{mhiggslong,mhiggsAEC}). In the MSSM it
corresponds to the case where all 
superpartners are heavy, i.e.\ the decoupling region of the MSSM.
The current 68~and~95\%~C.L.\ experimental results 
for $\mt$~\cite{mt1709},
\BE
\mt^{\rm exp} = 170.9 \pm 1.8 \gev ,
\EE
and $\MW$ are indicated in the plot. As can be seen from 
\reffi{fig:MWMTtoday}, the current experimental 68\%~C.L.\ region for 
$\mt$ and $\MW$ exhibits a slight preference of the MSSM over the SM;
only at the 95\%~C.L.\ the experimental values enter the SM parameter
space. This example indicates that the experimental measurement of $\MW$
in combination with $\mt$ 
prefers within the SM a relatively small value of $\MHSM$, or with the
MSSM not too heavy SUSY mass scales.

\section{The anomalous magnetic moment of the muon}
\label{sec:g-2}

Another important EWPO which is important in the context of precision tests
of the electroweak theory is the anomalous magnetic moment of the muon, 
$\amu \equiv (g-2)_\mu/2$. For the interpretation of the $\amu$ results
in the context of a NPM the current status of the comparison of the SM
prediction with the  experimental result is crucial, see
\citeres{g-2review,g-2review2,g-2reviewDS,g-2reviewMRR,g-2reviewFJ}
for reviews. 
$\amu$ is related to the 
photon--muon vertex function $\Gamma_{\mu\bar\mu A^\rho}$ as follows:
\BEA
\label{covdecomp}
&& \bar u(p')\Ga_{\mu\bar\mu A^\rho}(p,-p',q) u(p) \non \\
&=& \bar u(p')\big[\ga_\rho F_V(q^2) \\ 
&& + (p+p')_\rho F_M(q^2) + \ldots\big] u(p), \non \\
\amu & = & -2m_\mu F_M(0).
\EEA

The SM prediction for the anomalous magnetic moment of 
the muon depends on the evaluation of QED contributions (see
\citeres{Kinoshita,g-2QEDmassdep} for recent updates), the
hadronic vacuum polarization and light-by-light (LBL) contributions. The
former  have been evaluated
in~\citeres{g-2reviewFJ,g-2HMNT2,DDDD} 
and the latter in~\citeres{LBL,LBLnew,LBLnew2}. 
The evaluations of the 
hadronic vacuum polarization contributions using $e^+ e^-$ and $\tau$ 
decay data give somewhat different results. In view of the fact that
recent $e^+ e^-$ measurements tend to confirm earlier results, whereas
the correspondence between previous $\tau$ data and preliminary data
from BELLE is not so clear, and also in view of the additional
uncertainties associated with  
the isospin transformation from $\tau$ decay, nowadays the
$\tau$~results are usually discarded. This gives an estimate
based on $e^+e^-$ data~\cite{DDDD}:
\BEA
\amutheo &=& 
(11\, 659\, 180.5 \pm 4.4_{\rm had} \pm 3.5_{\rm LBL} \non \\
&&  \pm 0.2_{\rm QED+EW}) \times 10^{-10},
\label{eq:amutheo}
\EEA
where the source of each error is labeled. We note that the new $e^+e^-$
data sets that have recently been published in~\citeres{KLOE,CMD2,SND} have
been partially included in the updated estimate of $(g - 2)_\mu$, see
also \citere{KLOEeps}.

The SM prediction is to be compared with
the final result of the Brookhaven $(g-2)_\mu$ experiment 
E821~\cite{g-2exp}, namely:
\BE
\amuexp = (11\, 659\, 208.0 \pm 6.3) \times 10^{-10},
\label{eq:amuexp}
\EE
leading to an estimated discrepancy~\cite{DDDD,g-2SEtalk}
\BE
\amuexp-\amutheo = (27.5 \pm 8.4) \times 10^{-10},
\label{delamu}
\EE
equivalent to a 3.3-$\sigma$ effect%
\footnote{Three other recent evaluations yield slightly different
  numbers~\cite{g-2reviewMRR,g-2reviewFJ,g-2HMNT2}, 
  but similar discrepancies with the SM prediction.}%
.~While it would be premature to regard this deviation as a firm
evidence for new physics, it should be noted that this more than
$3\,\si$ effect has now firmly been established.

Taking the MSSM as an example to explain the $3.3\,\si$ effect,
the one-loop (and higher-order corrections) have to be evaluated.
The complete one-loop contribution to $\amu$ 
can be devided into
contributions from diagrams with a smuon-neutralino loop and with a
sneutrino-chargino loop, see \reffi{fig:g-21L}, leading to
\BE
\De\amuSUoL = \De\amu^{\tilde \chi^\pm \tilde \nu_\mu} +
               \De\amu^{\tilde \chi^0 \tilde \mu} .
\end{equation}

\setlength{\unitlength}{0.23mm}
\begin{figure}[htb!]
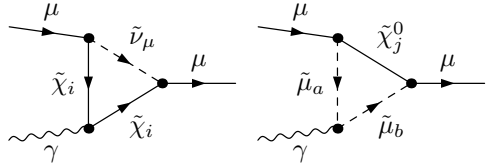

\vspace{-2em}
\begin{center}
\begin{feynartspicture}(288,168)(2,1)

\FADiagram{}
\FAProp(0.,15.)(7.,14.)(0.,){/Straight}{1}
\FALabel(3.7192,15.5544)[b]{$\mu$}
\FAProp(0.,5.)(7.,6.)(0.,){/Sine}{0}
\FALabel(3.7192,4.44558)[t]{$\gamma$}
\FAProp(20.,10.)(13.5,10.)(0.,){/Straight}{-1}
\FALabel(16.75,11.07)[b]{$\mu$}
\FAProp(7.,14.)(7.,6.)(0.,){/Straight}{1}
\FALabel(5.93,10.)[r]{$\tilde \chi_i$}
\FAProp(7.,14.)(13.5,10.)(0.,){/ScalarDash}{1}
\FALabel(10.5824,12.8401)[bl]{$\tilde \nu_\mu$}
\FAProp(7.,6.)(13.5,10.)(0.,){/Straight}{1}
\FALabel(10.5824,7.15993)[tl]{$\tilde \chi_i$}
\FAVert(7.,14.){0}
\FAVert(7.,6.){0}
\FAVert(13.5,10.){0}

\FADiagram{}
\FAProp(0.,15.)(7.,14.)(0.,){/Straight}{1}
\FALabel(3.7192,15.5544)[b]{$\mu$}
\FAProp(0.,5.)(7.,6.)(0.,){/Sine}{0}
\FALabel(3.7192,4.44558)[t]{$\gamma$}
\FAProp(20.,10.)(13.5,10.)(0.,){/Straight}{-1}
\FALabel(16.75,11.07)[b]{$\mu$}
\FAProp(7.,14.)(7.,6.)(0.,){/ScalarDash}{1}
\FALabel(5.93,10.)[r]{$\tilde \mu_a$}
\FAProp(7.,14.)(13.5,10.)(0.,){/Straight}{0}
\FALabel(10.4513,12.6272)[bl]{$\tilde \chi_j^0$}
\FAProp(7.,6.)(13.5,10.)(0.,){/ScalarDash}{1}
\FALabel(10.5824,7.15993)[tl]{$\tilde \mu_b$}
\FAVert(7.,14.){0}
\FAVert(7.,6.){0}
\FAVert(13.5,10.){0}

\end{feynartspicture}

\end{center}
\vspace{-4em}
\caption[]{
The generic one-loop diagrams for the MSSM contribution to $\amu$:
diagram with a sneutrino-chargino loop (left) and the diagram with a
smuon-neutralino loop (right).
}
\label{fig:g-21L}
\end{figure}

The coupling of an external muon to the SUSY particles is enhanced by
$\tb$, which can range from $\sim 2$ to $\sim 60$. This can lead to a
strong enhancement of the MSSM one-loop diagrams in comparison with the
corresponding SM one-loop electroweak diagrams, despite the fact that
the masses of the SM particles involved are lighter than the SUSY mass
scales. 
The full one-loop expression can be found in~\citere{g-2MSSMf1l}, see
\citere{g-2early} for earlier evaluations. If all SUSY
mass scales are set to a common value, 
$\msusy = m_{\cha{}} = m_{\neu{}} = m_{\Smu} = m_{\Sneum}$, 
the result is given by
\BEA
\amu^{\SU,{\rm 1L}} &=& 13 \times 10^{-10}
             \KL \frac{100 \gev}{\msusy} \KR^2 \non \\ 
&& \times \tb\;  {\rm  sign}(\mu) ~.
\label{susy1loop}
\EEA
Obviously, supersymmetric effects can easily account for a
$(20 \ldots 40) \times 10^{-10}$ deviation, if $\mu$ is positive and
$\msusy$ lies roughly between 100 GeV (for small $\tb$) and
600 GeV (for large $\tb$). On the other hand, demanding that SUSY
fulfills \refeq{delamu} at the two or three $\si$ level, \refeq{susy1loop}
shows that the $(g-2)_\mu$ measurement places strong bounds on the
supersymmetric parameter space. 

In addition to the full one-loop contributions, the leading QED
two-loop corrections have also been
evaluated~\cite{g-2MSSMlog2l}. Further corrections at the two-loop
level have been obtained~\cite{g-2FSf,g-2CNH}, 
leading to corrections to the one-loop result that are $\lsim 10\%$.
These corrections are taken into account in the examples shown below.

\begin{figure}[htb!]
\begin{center}
\includegraphics[width=.45\textwidth,height=.35\textwidth] 
                {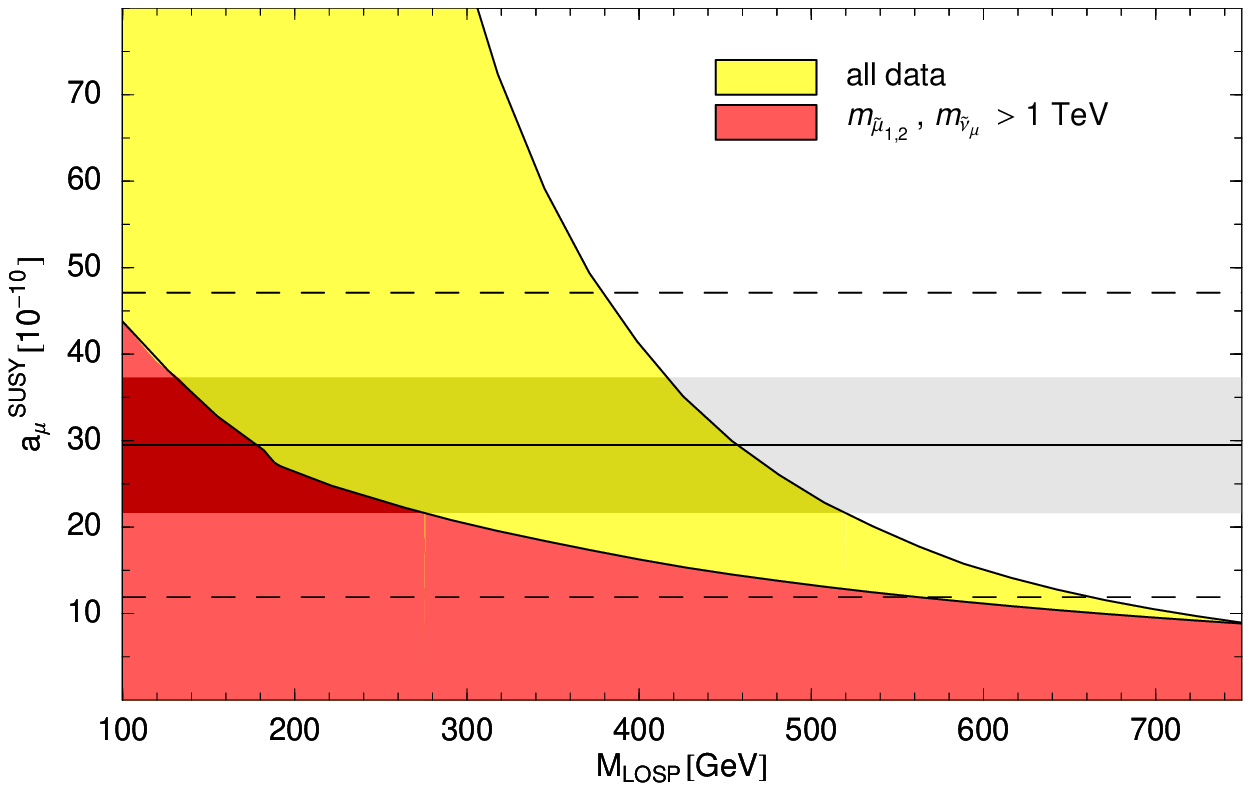}\\[1em]
\includegraphics[width=.45\textwidth,height=.35\textwidth] 
                {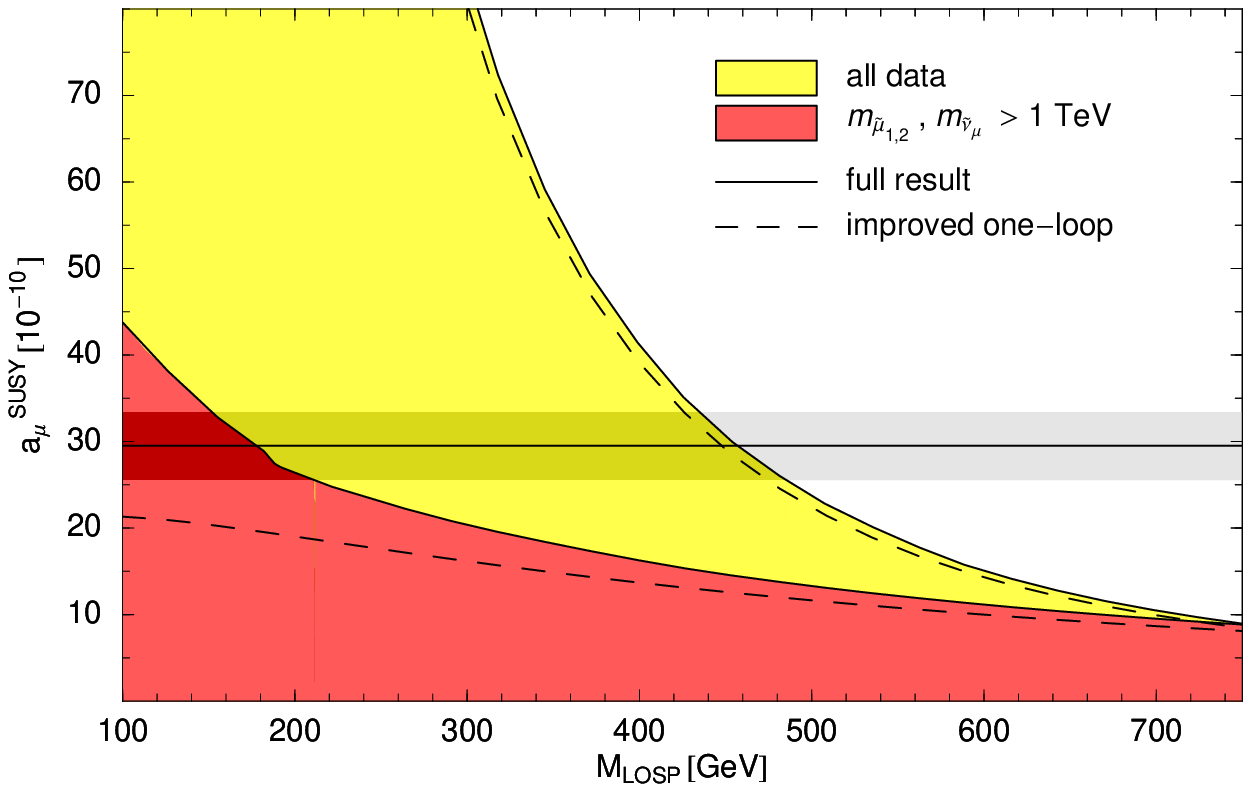}
\vspace{-1em}
\caption{%
Scan over the MSSM parameter space for $\amu^{\SU}$, including all
available one- and two-loop contributions, as a function of
the lightest observables particle (the lightest $\Smu$ or $\neu{2}$ or
$\cha{1}$)~\cite{g-2reviewDS}. The light (yellow) shaded region is all
data, the dark 
(red) shaded region has all second generation sleptons heavier than 
$1 \tev$. The upper plot shows the current experimental deviation at
the one (shaded stip) or two (dashed lines) $\si$ level. The
lower plot shows the full result for $\amu^{\SU}$ (full area) and the
one-loop result only (dashed lines). Here the shaded strip
corresponds to an anticipated future accuracy of $4 \times 10^{-10}$
(see text).
}
\label{fig:g-2Scan}
\end{center}
\vspace{-1em}
\end{figure}

Concerning other NPM, the generic size of the new contribution to $\amu$ is
roughly given in in terms of the NPM mass scale 
$M_{\rm NPM}$~\cite{g-2reviewDS}, 
\BE
\amu^{\rm NPM} \sim 1 \times 10^{-10} 
               \KL \frac{300 \gev}{M_{\rm NPM}} \KR^2 .
\EE
Thus, the generic NPM contribution is usually too small to explain the
$3.3\,\si$ effect in \refeq{delamu}. The advantanges of SUSY are the
$\tb$ enhancement of the muon coupling to SUSY particles and the fact
that relatively light SUSY particles with masses $\gsim 100 \gev$ are
experimentally allowed. 

In \reffi{fig:g-2Scan} we show the results of an MSSM parameter scan
for $\amu$, including all
available one- and two-loop contributions, as a function of
the lightest observables particle (the lightest $\Smu$ or $\neu{2}$ or
$\cha{1}$)~\cite{g-2reviewDS}. The light (yellow) shaded region is all
data, the dark (red) shaded region has all second generation sleptons
heavier than $1 \tev$. In the lower plot of \reffi{fig:g-2Scan} the
prediction of the one-loop result only is indicated by the
dashed lines. It can be clearly seen that making the smuons and
charginos/neutralinos heavy suppresses the one-loop diagrams
shown in \reffi{fig:g-21L}. In this case the two-loop contribution
become important~\cite{g-2FSf,g-2CNH}. The upper plot shows the
current one (shaded stip) or two (dashed lines) $\si$ results
according to \refeq{delamu}. It can be clearly seen that demanding
agreement of the MSSM contribution with the current experimental
result imposes strong restrictions on the parameter space.

A new $(g-2)_\mu$ experiment has been proposed, see \citere{g-2expNew}
and references therein. Together with further improvement on the
theory side, the error of $\amuexp-\amutheo$ could be decreased to the
level of $4 \times 10^{-10}$~\cite{g-2reviewDS,g-2expNew}. The effect
of this anticipated future precision can be seen in the lower plot of
\reffi{fig:g-2Scan}, assuming the current central deviation. The
restrictions on the MSSM parameter space would become very strong. The
case with heavy smuons and charginos/neutralinos could only be
realized using the SUSY prediction at the two-loop
level~\cite{g-2FSf,g-2CNH}.


\section{Electric Dipole Moments}
\label{sec:edm}

A different way for probing NPM is via their contribution to EDMs of
heavy quarks, of the electron and the neutron or neutral
atoms. Some present limits are summarized in \refta{tab:edmtoday}, see
\citere{EDMtoday} for a review. Improvements of the sensitivities of
\order{10^1-10^2} can be expected from ongoing and future experiments,
see \citere{ShufangPhysRept} (and references therein).

\begin{table}[htb!]
\renewcommand{\arraystretch}{1.2}
\begin{center}
{\small
\begin{tabular}{|c|c|c|} \hline
System & limit & group \\ \hline\hline
$e^-$  & $1.6 \times 10^{-27}$ (90\% C.L.) & Berkely \\ \hline
$n$    & $2.9 \times 10^{-26}$ (90\% C.L.) & ILL     \\ \hline
$\mbox{}^{199}$Hg & $2.1 \times 10^{-28}$  (95\% C.L.) & Seattle \\ 
\hline \hline
\end{tabular}
}
\vspace{-2em}
\caption{Present bounds for EDMs~\cite{EDMe,EDMn,EDMhg}.}
\label{tab:edmtoday}
\end{center}
\vspace{-1em}
\end{table}

While SM contributions start only at the three-loop level~\cite{EDMSM}, 
due new complex parameters NPM can contribute already at one-loop
order~\cite{EDMrevNPM}. 
Taking the MSSM with complex parameters (cMSSM) as a specific example,
the respective calculations can 
be found for heavy quarks in \citere{EDMDoink}, for the electron and 
the neutron in \citeres{EDMrev2,EDMPilaftsis} and references therein.
Recent reviews concerning the
EDMs in the cMSSM are given in \citeres{EDMrev2,EDMrev1,EDMrev3}. 

\begin{figure}[htb!]
\begin{center}
\includegraphics[width=.45\textwidth,height=.25\textwidth] 
                {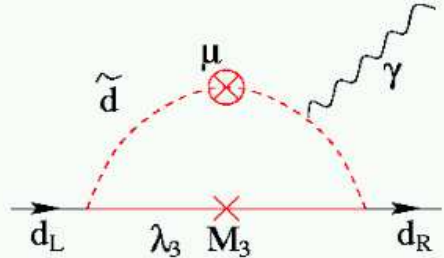}
\vspace{-1em}
\caption{%
Generic SUSY diagram contributing to the EDM of the neutron.
}
\label{fig:susyEDM}
\end{center}
\vspace{-1em}
\end{figure}

A generic SUSY diagram is given in \reffi{fig:susyEDM} yielding a
contribution to the EDM of the neutron, $d_n$, as~\cite{EDMefn}
\BE
\frac{d_n}{m_d} \sim \frac{1}{16\pi^2}\,
                         \frac{\mu\,\mgl}{\msusy}\,
                         \sin\theta_\mu ,
\EE
where $m_d$ is the mass of the down quark, $\mgl$ denotes the gluino
mass, and $\mu$ is the Higgs mixing parameter with its phase
$\theta_\mu$. 

\noindent
Also the
leading two-loop corrections for the electron and neutron EDMs are
available~\cite{EDMPilaftsis,EDM2l}. Large phases in the first two
generations of (s)fermions 
can only be accomodated if these generations are assumed to be very
heavy~\cite{EDMheavy} or large cancellations occur~\cite{EDMmiracle},
see however the discussion in \citere{EDMrev1}.
EDMs thus place already strong bounds on the size of the complex phases
of the cMSSM (see e.g.\ \citere{plehnix}) and have to be taken into
account in any related phenomenological analysis.


\section{EWPO in the SM}

Within the SM the EWPO have been used to constrain the last unknown
parameter of the model, the Higgs-boson mass $\MHSM$. 
Originally the EWPO comprise over thousand measurements of ``realistic
observables'' (with partically correlated uncertainties) such as cross
sections, asymmetries, branching ratios etc. This huge set is reduced to
17 so-called ``pseudo observables'' by the LEP~\cite{LEPEWWG} and
Tevatron~\cite{TEVEWWG} Electroweak working groups. 
The ``pseudo observables'' (again called EWPO in the following) comprise
the $W$~boson mass $\MW$ (see \refse{sec:mw}), the width of the
$W$~boson, $\Ga_W$, as well as various $Z$~pole observables: the
effective weak mixing angle, $\sweff$, $Z$~decay widths to SM fermions, 
$\Ga(Z \to f \bar f)$, the invisible and total width, $\Ga_{\rm inv}$ and
$\Ga_Z$, forward-backward and left-right asymmetries, $A_{\rm FB}^f$ and
$A_{\rm LR}^f$, and the total hadronic cross section, $\si^0_{\rm had}$.
The $Z$~pole results including their combination are
final~\cite{lepewwg}. Experimental progress from the Tevatron comes for
$\MW$ and $\mt$. (Also the error combination for $\MW$ and $\Ga_W$ from
the four LEP experiments has not been finalized yet due to not-yet-final
analyses on the color-reconnection effects.)

The EWPO that give the strongest constraints on $\MHSM$ are $\MW$, 
$A_{\rm FB}^b$ and $A_{\rm LR}^e$. The value of $\sweff$ is extracted
from a combination of various $A_{\rm FB}^f$ and $A_{\rm LR}^f$, where 
$A_{\rm FB}^b$ and $A_{\rm LR}^e$ give the dominant contribution.

\begin{figure}[htb!]
\vspace{-1em}
\begin{center}
\includegraphics[width=.45\textwidth,height=.5\textwidth] 
                {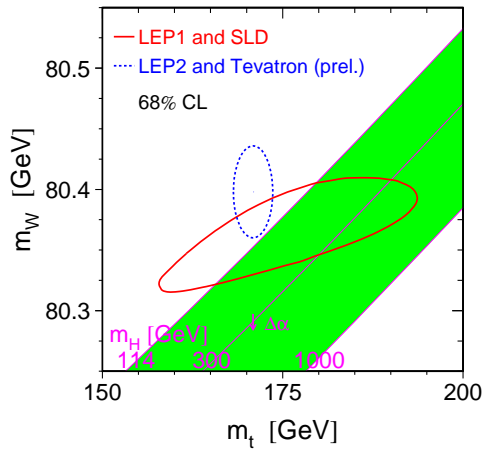}
\vspace{-2em}
\caption{%
Prediction for $\MW$ in the SM as a function of $\mt$ for the range 
$\MHSM = 114 \gev \ldots 1000 \gev$~\cite{LEPEWWG}. The prediction is
compared with  
the present experimental results for $\MW$ and $\mt$ as well as with the
indirect constraints obtained from EWPO.
}
\label{fig:MWMTSM}
\end{center}
\vspace{-1em}
\end{figure}

The one-loop contributions to $\De r$ (i.e.\ to $\MW$, see
\refeq{eq:delr}) can be decomposed as follows~\cite{sirlin},
\BE
\De r_{1-{\rm loop}} = \De\al - \frac{\cw^2}{\sw^2}\De\rho 
                     + \De r_{\rm rem}(\MHSM) .
\label{eq:deltar1l}
\EE
The first term, $\De\al$ contains large logarithmic contributions as
$\log(\MZ/m_f)$ and amounts $\sim 6\%$. The second term contains the 
$\rho$~parameter~\cite{rho}, being $\De\rho \sim \mt^2$ (with 
$\cw^2 = \MW^2/\MZ^2$, $\sw^2 = 1 - \cw^2$). This term
amounts $\sim 3.3\%$. The final term in \refeq{eq:deltar1l} is
$\De r_{\rm rem} \sim \log(\MHSM/\MW)$, and with a size of $\sim 1\%$
correction yields the constraints on $\MHSM$. The fact that the leading
correction involving $\MHSM$ is logarithmic also applies to the other
EWPO. Starting from two-loop order, also terms $\sim (\MHSM/\MW)^2$
appear. The SM prediction of $\MW$ as a function of $\mt$ for the range
$\MHSM = 114 \gev \ldots 1000 \gev$ is shown as the dark shaded (green)
band in \reffi{fig:MWMTSM}~\cite{LEPEWWG}. The upper edge with 
$\MHSM = 114 \gev$ corresponds to the lower limit on $\MHSM$ obtained at
LEP~\cite{LEPHiggsSM}. The prediction is compared
with the direct experimental result (dotted/blue ellipse) and with the
indirect results for $\MW$ and $\mt$ as obtained from EWPO (solid/red
ellipse). Consistent with \reffi{fig:MWMTtoday} the direct experimental
result at the 68\%~C.L. does not enter the SM prediction, i.e.\ low 
SM Higgs boson masses, $\MHSM \sim 44 \gev$~\cite{gruenewaldpriv}, are
preferred by the measurement of $\MW$ and $\mt$.

\begin{figure}[htb!]
\begin{center}
\includegraphics[width=.45\textwidth,height=.5\textwidth] 
                {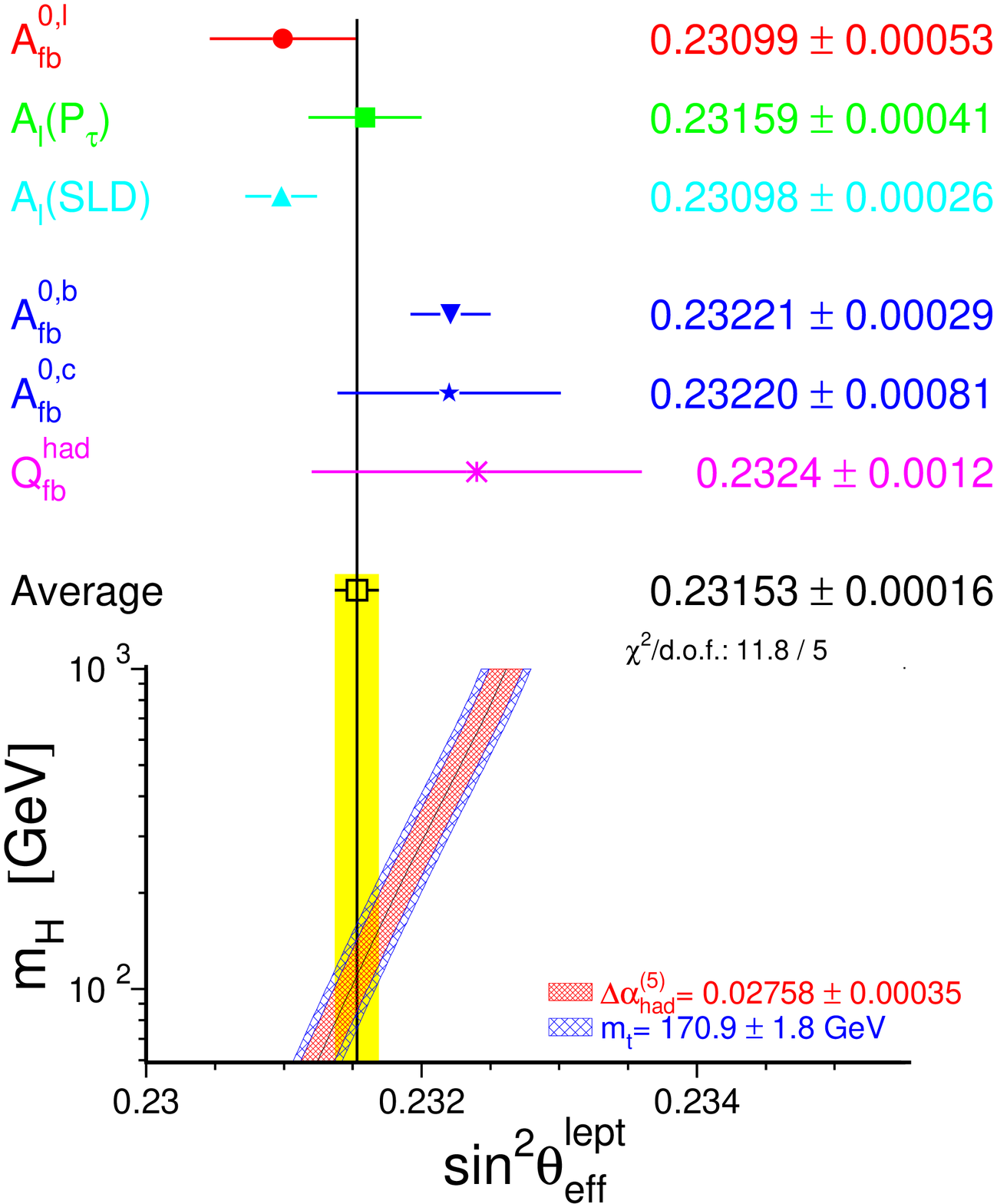}
\vspace{-1em}
\caption{%
Prediction for $\sweff$ in the SM as a function of $\MHSM$ for 
$\mt = 170.9 \pm 1.8 \gev$ and 
$\De\al_{\rm had}^5 = 0.02758 \pm 0.00035$~\cite{gruenewald07}. The
prediction is compared with  
the present experimental results for $\sweff$ as averaged over several
individual measurements.
}
\label{fig:sw2effSM}
\end{center}
\vspace{-1em}
\end{figure}

The effective weak mixing angle is evaluated from various asymmetries
and other EWPO as shown in \reffi{fig:sw2effSM}~\cite{gruenewald07}. The
average determination yields $\sweff = 0.23153 \pm 0.00016$ with a
$\chi^2/{\rm d.o.f}$ of $11.8/5$, corresponding to a probability of
$3.7\%$~\cite{gruenewald07}. The large $\chi^2$ is driven by the two
single most precise measurements, $A_{\rm LR}^e$ by SLD and 
$A_{\rm FB}^b$ by LEP, where the earlier (latter) one prefers a 
value of $\MHSM \sim 32 (437) \gev$~\cite{gruenewaldpriv}. 
The two measurements differ by more than $3\,\si$.
The averaged value of $\sweff$, as shown in \reffi{fig:sw2effSM},
prefers $\MHSM \sim 110 \gev$~\cite{gruenewaldpriv}. 

\begin{figure}[htb!]
\begin{center}
\includegraphics[width=.45\textwidth,height=.4\textwidth] 
                {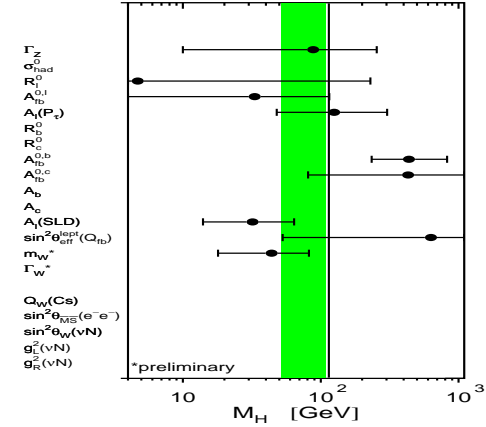}
\vspace{-1em}
\caption{%
Indirect constrains on $\MHSM$ from various EWPO. Shown are the central
values and the one~$\si$ errors~\cite{LEPEWWG}. 
The dark shaded (green) vertical band indicates the combination of the
various single measurements in the $1\,\si$ range. The vertical line shows
the lower bound of $\MHSM \ge 114.4 \gev$ obtained at
LEP~\cite{LEPHiggsSM}.
}
\label{fig:MHSM}
\end{center}
\vspace{-1em}
\end{figure}

The indirect $\MHSM$ determination for several individual EWPO is given
in \reffi{fig:MHSM}. Shown are the central
values of $\MHSM$ and the one~$\si$ errors~\cite{LEPEWWG}. 
The dark shaded (green) vertical band indicates the combination of the
various single measurements in the $1\,\si$ range. The vertical line shows
the lower LEP bound for $\MHSM$~\cite{LEPHiggsSM}.
It can be seen that $\MW$, $A_{\rm LR}^e$ and $A_{\rm FB}^b$ give the
most precise indirect $\MHSM$ determination, where only the latter one
pulls the preferred $\MHSM$ value up, yielding a averaged value
of~\cite{LEPEWWG} 
\BE
\MHSM = 76^{+33}_{26} \gev~,
\label{MHSMfit}
\EE
still compatible with the direct LEP bound of~\cite{LEPHiggsSM}
\BE
\MHSM \ge 114.4 \gev \mbox{~at~} 95\% \mbox{~C.L.}
\label{MHSMlimit}
\EE
Thus, the measurement of $A_{\rm FB}^b$ prevents the SM from being
incompatible with the direct bound and the indirect constraints on
$\MHSM$. 

\begin{figure}[htb!]
\vspace{-1em}
\begin{center}
\includegraphics[width=.45\textwidth,height=.50\textwidth] 
                {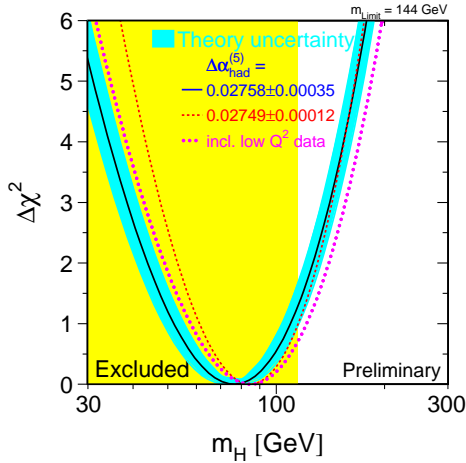}
\vspace{-2em}
\caption{%
$\De\chi^2$ curve derived from all EWPO measured at LEP, SLD, CDF and
D0, as a function of $\MHSM$, assuming the SM to be the correct theory
of nature~\cite{LEPEWWG}. 
}
\label{fig:blueband}
\end{center}
\vspace{-1em}
\end{figure}

Finally, in \reffi{fig:blueband}~\cite{LEPEWWG} we show the result for
the global fit to $\MHSM$ including all EWPO. $\De\chi^2$ is shown as a
function of $\MHSM$, yielding \refeq{MHSMfit} as best fit with an upper
limit of $144 \gev$ at 95\%~C.L. This value increases to $182 \gev$ if
the direct LEP bound of \refeq{MHSMlimit} is included in the fit. The
theory (intrinsic) uncertainty in the SM calculations (as evaluated with 
{\tt TOPAZ0}~\cite{topaz0} and {\tt ZFITTER}~\cite{zfitter}) are
represented by the thickness of the blue band. The width of the parabola
itself, on the other hand, is determined by the experimental precision of
the measurements of the EWPO and the input parameters.

The current and anticipated future experimental uncertainties for
$\sweff$, $\MW$ and $\mt$ are summarized in \refta{tab:POfuture}. Also
shown is the relative precision of the 
indirect determination of $\MHSM$~\cite{gruenewald07}.
Each column represents the combined results of all detectors and
channels at a given collider, taking into account correlated
systematic uncertainties, see \citeres{blueband,gigaz,moenig,mwgigaz}
for details. The indirect $\MHSM$ determination has to be compared with
the (possible) direct measurement at the LHC~\cite{atlas,cms} and the
ILC~\cite{teslatdr,orangebook,acfarep,Snowmass05Higgs}, 
\BEA
\label{deltaMHLHC}
\de\MHSM\mbox{}^{\rm ,exp,LHC} &\approx& 200 \mev ,\\
\label{deltaMHILC}
\de\MHSM\mbox{}^{\rm ,exp,ILC} &\approx& 50 \mev .
\EEA

\begin{table}[htb!]
\renewcommand{\arraystretch}{1.5}
\begin{center}
{\small
\begin{tabular}{|c||c|c|c|}
\cline{2-4} \multicolumn{1}{c||}{}
& now & Tevatron & LHC \\
\hline\hline
$\de\sweff(\times 10^5)$ & 16   & --- & 14--20 \\
\hline
$\de\MW$ [MeV]           & 25   &  20 & 15   \\
\hline
$\de\mt$ [GeV]           &  1.8 &  1.2 &  1.0 \\
\hline
$\de\MHSM/\MHSM$ [\%]     &  37 &     &  28 \\
\hline
\end{tabular}

\begin{tabular}{|c||c|c|}
\cline{2-3} \multicolumn{1}{c||}{}
& ~ILC~  & ILC with GigaZ \\
\hline\hline
$\de\sweff(\times 10^5)$ & ---  & 1.3  \\
\hline
$\de\MW$ [MeV]           & 10   & 7      \\
\hline
$\de\mt$ [GeV]           &  0.2 & 0.1   \\
\hline
$\de\MHSM/\MHSM$ [\%]    &      & 16 \\
\hline
\end{tabular}
}
\end{center}
\renewcommand{\arraystretch}{1}
\vspace{-1em}
\caption{
Current and anticipated future experimental uncertainties for
$\sweff$, $\MW$ and $\mt$. Also shown is the relative precision of the
indirect determination of $\MHSM$~\cite{gruenewald07}.
Each column represents the combined results of all detectors and
channels at a given collider, taking into account correlated
systematic uncertainties, see \citeres{blueband,gigaz,moenig,mwgigaz}
for details. 
}
\label{tab:POfuture}
\vspace{1em}
\end{table}

\noindent
This comparison will shed light on the basic theoretical components for
generating the masses of the fundamental particles.  
On the other hand, an observed inconsistency would be a clear
indication for the existence of a new physics scale.


\newpage
\section{EWPO in the MSSM}

As compared to the SM there are new additional contributions to EWPO in
the MSSM that can be sizable:
\begin{enumerate}
\item
While in the SM the leading corrections to e.g.\ the $\rho$~parameter
(i.e.\ to gauge boson self-energies)
arise from $t/b$~loops, in the MSSM large corrections can arise from
$\Stop/\Sbot$~loops ($V = Z, W^\pm$):

\unitlength=0.7cm%
\begin{feynartspicture}(4,4)(1,1)
\FADiagram{}
\FAVert(5,10){0}
\FAVert(15,10){0}
\FAProp(0,10)(5,10)(0.,){/Sine}{0}
\FAProp(15,10)(20,10)(0.,){/Sine}{0}
\FALabel(2,12)[r]{$V$}
\FALabel(18,12)[l]{$V$}
{
\FAProp(5,10)(15,10)(1.,){/Straight}{0}
\FAProp(15,10)(5,10)(1.,){/Straight}{0}
\FALabel(8.5,15.5)[b]{\ ${t}$,${b}$}}
\end{feynartspicture}
\begin{feynartspicture}(4,4)(1,1)
\FADiagram{}
\FAVert(5,10){0}
\FAVert(15,10){0}
\FAProp(0,10)(5,10)(0.,){/Sine}{0}
\FAProp(15,10)(20,10)(0.,){/Sine}{0}
\FALabel(2,12)[r]{$V$}
\FALabel(18,12)[l]{$V$}
{
\FAProp(5,10)(15,10)(1.,){/ScalarDash}{0}
\FAProp(15,10)(5,10)(1.,){/ScalarDash}{0}
\FALabel(8.5,15.5)[b]{\ $\tilde{t}$,$\tilde{b}$}}
\end{feynartspicture}
\vspace{-3em}

\item
New $\cp$-violating effects can arise from new complex parameters, see
\refse{sec:edm}. 

\item
Yukawa corrections $\sim \mt^4 \log\KL\frac{\mste\mstz}{\mt^2}\KR$
can give large contributions.

\item
Corrections from the $b/\Sbot$~sector are enhanced by $\tb$ and can
become sizable, see also \refse{sec:g-2}.

\item
In general SUSY corrections are relevant if the new mass scales are
(relatively) small. On the other hand, 
non-decoupling SUSY effects \mbox{$\sim \log \frac{\msusy}{\MW}$} can become
important for large values of $\msusy$.

\end{enumerate}

The example of the $W$~boson mass has been discussed in
\refse{sec:mw}. In the same spirit also the $\sweff$ has been evaluated
in the MSSM and compared to the SM prediction~\cite{ZOpope}. A parameter
scan similar to the one shown in \reffi{fig:MWMTtoday} reveals no preference
for either model. This result, as the not too low best-fit value for
$\MHSM$ is largely driven by the measurement of $A_{\rm FB}^b$, while
$A_{\rm LR}^e$ has a clear preference for the MSSM prediction.

\medskip
Another EWPO in the MSSM is the mass of the lightest Higgs boson,
$\Mh$. In the MSSM two Higgs doublets are required,
resulting in five physical Higgs bosons: the light and heavy $\cp$-even $h$
and $H$, the $\cp$-odd $A$, and the charged Higgs bosons $H^\pm$.
The Higgs sector of the MSSM can be expressed at lowest
order in terms of $\MZ$, $\MA$ and $\tb$. All other masses and
mixing angles can therefore be predicted. 
At the tree-level this leads to the prediction of $\Mh^{\rm tree} \le \MZ$.
However, the tree-level bound on $\Mh$, being obtained from the gauge
couplings, receives 
large corrections from SUSY-breaking effects in the Yukawa sector of the 
theory. The leading one-loop correction is proportional to $\mt^4$.
The leading logarithmic one-loop term (for vanishing mixing
between the scalar top quarks) reads~\cite{mhiggs1l} 
\BE
\De \Mh^2 = \frac{3 \GF \mt^4}{\wz\, \pi^2\,\SQb}
          \log \KL \frac{\mste \mstz}{\mt^2} \KR~.
\label{deltamhmt4}
\end{equation}
Corrections of this kind have drastic effects on the predicted value of
$\Mh$ and many other observables in the MSSM Higgs sector. The one-loop
corrections can shift $\Mh$ by 50--100\%. 
In this way the MSSM Higgs sector, and especially $\Mh$, depend
sesitively on the other MSSM paramters; $\Mh$ will be the most powerful
precision observable in the MSSM. 

The status of higher-order corrections to the masses (and the mixing) in
the Higgs sector of the MSSM%
\footnote{We concentrate here on the case with real paramters. For
  complex parameters see \citeres{mhcMSSMlong,mhcMSSM2L} and references
  therein.}
is quite advanced. The complete one-loop result within the MSSM is
known~\cite{mhiggs1l,mhiggsf1lA,mhiggsf1lB,mhiggsf1lC}.
The by far dominant
one-loop contribution is the \order{\alt} term due to top and stop
loops ($\alt \equiv h_t^2 / (4 \pi)$, $h_t$ being the
top-quark Yukawa coupling). The computation of the two-loop corrections
has meanwhile reached a stage where all the presumably dominant
contributions are 
available~\cite{mhiggslong,mhiggsletter,mhiggslle,mhiggsFD2,bse,mhiggsEP03b,mhiggsEP124b,mhiggsRG1a},
see \citeres{mhiggsAEC,PomssmRep} for reviews.
In particular, the \order{\alt\als}, \order{\alt^2}, \order{\alb\als},
\order{\alt\alb} and \order{\alb^2} contributions to the self-energies
are known for vanishing external momenta.  For the (s)bottom
corrections, which are mainly relevant for large values of $\tb$,
an all-order resummation of the $\tb$-enhanced term of
\order{\alb(\als\tb)^n} is performed~\cite{deltamb2,deltamb2b3}. 
The remaining theoretical uncertainty on the lightest $\cp$-even Higgs
boson mass has been estimated to be below 
$\sim 3 \gev$~\cite{mhiggsAEC,PomssmRep,mhiggsWN}.  
The above calculations have been implemented into public 
codes. The program 
{\tt FeynHiggs}~\cite{mhiggslong,mhiggsAEC,mhcMSSMlong,feynhiggs}
is based on the results obtained in the Feynman-diagrammatic (FD)
approach and includes all the above corrections.
The code {\tt CPsuperH}~\cite{cpsh} is based on the renormalization group (RG)
improved effective potential approach.
Most recently a full two-loop effective potential calculation
(including even the momentum dependence for the leading
pieces and the leading three-loop corrections) has been
published~\cite{mhiggsEP5}. However, no computer code is publicly
available. 

While a precise knowledge of $\mt$ is important for $\MW$, $\sweff$,
\ldots, it is {\em crucial} for $\Mh$, see also \citere{deltamt}. Due
to the strong dependence of $\Mh$ on $\mt$, see \refeq{deltamhmt4}, by
numerical coincidence 
\BE
\de\mt^{\rm exp}/\de\Mh^{\rm theo} \approx 1
\EE
holds~\cite{tbexcl}. Thus already the LHC precision for $\Mh$, 
\refeq{deltaMHLHC}, requires the ILC precision for $\mt$, see
\refta{tab:POfuture}. (More examples of such LHC/ILC interplay can be
found in \citere{lhcilc}.)


\section{EWPO in the CMSSM}

In order to achieve a simplification of the plethora of soft
SUSY-breaking parameters appearing in the general MSSM, 
one assumption that is frequently employed is
that (at least some of) the soft SUSY-breaking parameters are universal
at some high input scale, before renormalization. 
One model based on this simplification is the 
constrained MSSM (CMSSM), in which all the soft SUSY-breaking scalar
masses $m_0$ are assumed to be universal at the GUT scale, as are the
soft SUSY-breaking gaugino masses $m_{1/2}$ and trilinear couplings
$A_0$. Further parameters are $\tb$ and the sign of the Higgs mixing
parameter $\mu$. 
Since the low-scale parameters in this scenario are derived from a small
set of input quantities, it is meaningful to combine various
experimental constraints. The EWPO can be supplemented
with $B$~physics observables (BPO) and astrophysical results such as the
cold dark matter (CDM) abundance.

\begin{figure}[htb!]
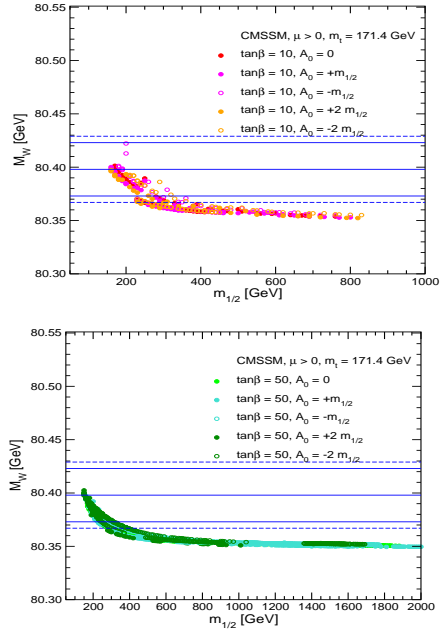

\begin{center}
\includegraphics[width=.4\textwidth,height=4cm]{ehow5.MW11a.1714.cl.eps}\\[3mm]
\includegraphics[width=.4\textwidth,height=4cm]{ehow5.MW11b.1714.cl.eps}
\vspace{-1em}
\caption{%
The CMSSM predictions for $\MW$ are shown as functions of $m_{1/2}$ along the 
WMAP strips for $\tb = 10$ (upper) and $\tb = 50$ (lower plot) for
various $A_0$ 
values~\cite{ehoww}. In each panel, the center (solid) line is the
present central experimental value, and the (solid) outer lines show the
current $\pm 1$-$\sigma$ range. 
The dashed lines correspond to the full error including also parametric
and intrinsic uncertainties.
}
\label{fig:MW}
\end{center}
\end{figure}

As an example we show the prediction for $\MW$ in the
CMSSM~\cite{ehoww}. The parameter points are chosen such that they
yield the correct value of the CDM density inferred
from WMAP and other data, namely~\cite{WMAP}
\BE 
0.094 < \Omega_{\rm CDM} h^2 < 0.129 .
\label{cdmexp}
\EE
The fact that the density is relatively well known
restricts the SUSY parameter space to a thin, fuzzy `WMAP
hypersurface'~\cite{WMAPstrips,wmapothers}, 
effectively reducing its dimensionality by one. The analysis has been
performed on `WMAP lines' in the \plane{m_{1/2}}{m_0}s for discrete
values of the other SUSY parameters: $\tb = 10, 50$ and
$A_0 = 0, \pm 1, \pm 2 \times m_{1/2}$.
In \reffi{fig:MW} the CMSSM prediction for $\MW$ is shown as a function of
$m_{1/2}$. The center (solid) line is the
present central experimental value, and the (solid) outer lines show the
current $\pm 1$-$\sigma$ range. 
The dashed lines correspond to the full error including also parametric
and intrinsic uncertainties.
One can see in that the variation with $A_0$ is
relatively weak for both values of $\tb$. The best results are obtained
for low $m_{1/2}$, while large values lead to a $\sim 1.5\,\si$
deviation (corresponding to the SM limit).

\begin{figure}[htb!]
\begin{center}
\includegraphics[width=.4\textwidth,height=4cm]
                {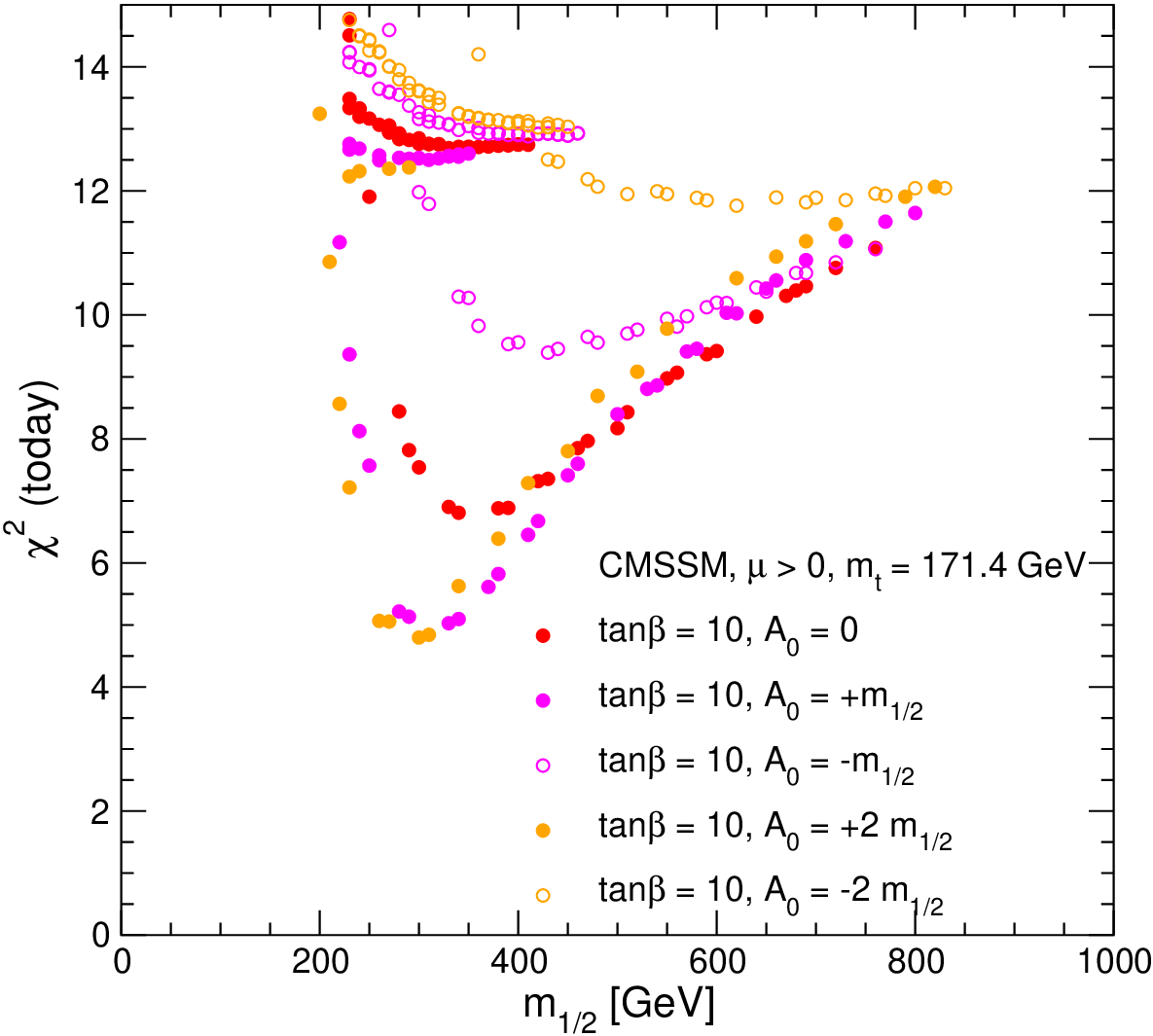}\\[3mm]
\includegraphics[width=.4\textwidth,height=4cm]
                {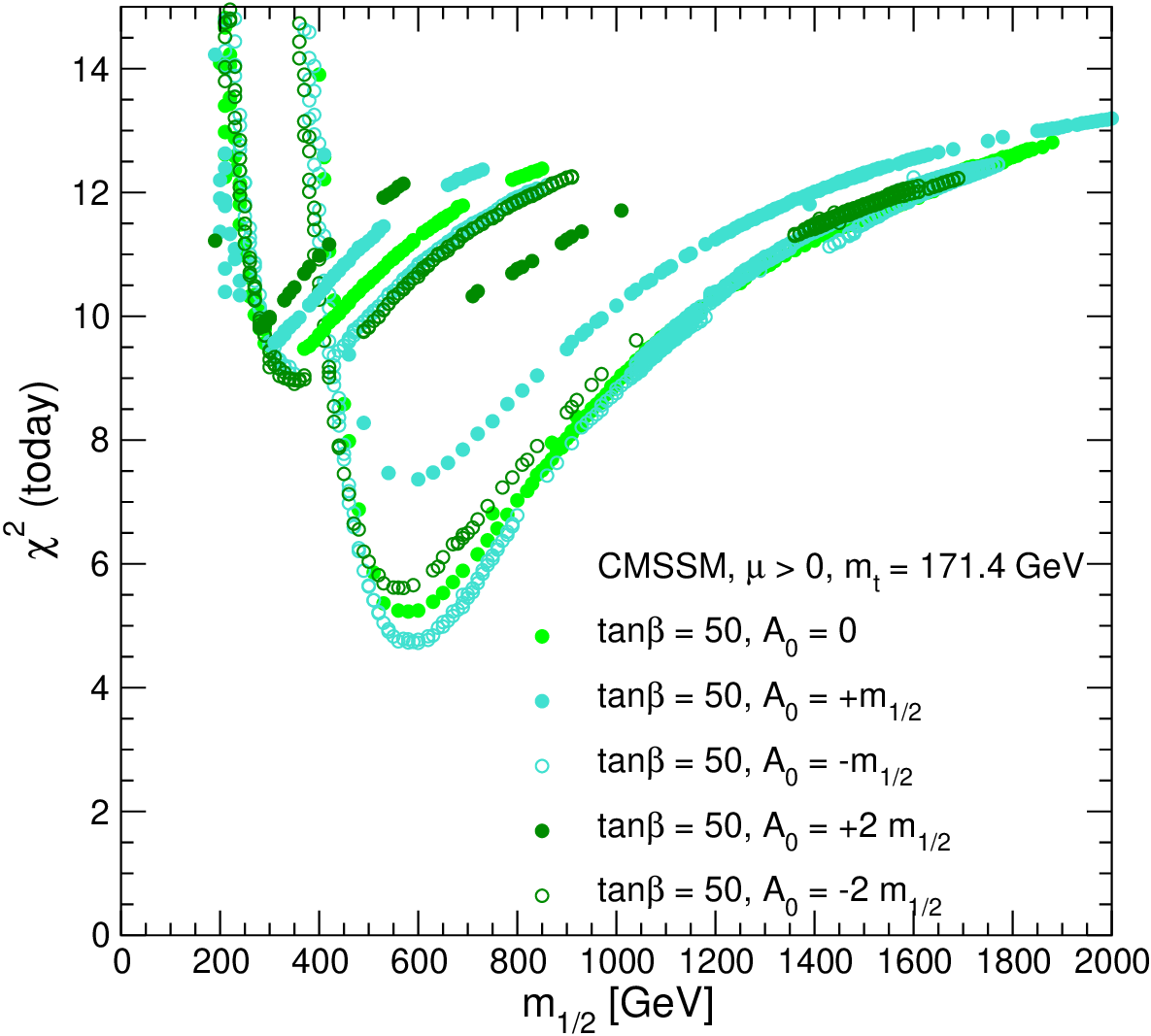}
\vspace{-1em}
\caption{%
The combined $\chi^2$~function for the 
EWPO $\MW$, $\sweff$, $\Ga_Z$, $(g - 2)_\mu$, $\Mh$, 
and the BPO
$\br(b \to s \ga)$, $\br(B_s \to \mu^+\mu^-)$, $\br(B_u \to \tau \nu_\tau)$
and $\De M_{B_s}$, evaluated in the CMSSM for $\tb = 10$ (upper) and
$\tb = 50$ (lower plot) for various discrete values of $A_0$~\cite{ehoww}.
}
\label{fig:CHI}
\end{center}
\end{figure}

In \citere{ehoww} five EWPO ($\MW$, $\sweff$, $\Ga_Z$, $(g-2)_\mu$ and
$\Mh$) and four BPO ($\br(b \to s \ga)$, $\br(B_s \to \mu^+\mu^-)$, 
$\br(B_u \to \tau \nu_\tau)$ and $\De M_{B_s}$) are used to perform a
$\chi^2$ analysis. For $\Mh$ and $\br(B_s \to \mu^+\mu^-)$ no
experimental obervation but the full experimental exclusion bounds (as
translated into $\chi^2$) have been used. Within the CMSSM the
lightest Higgs boson has SM-like production and decay
properties~\cite{asbs1,ehow1}, and the SM results~\cite{LEPHiggsSM}
can be used. The other parameters are
$\mt = 171.4 \pm 2.1 \gev$ and $\mb(\mb) = 4.25 \pm 0.11 \gev$, and 
$m_0$ is chosen to yield the central value of the cold dark matter
density indicated by WMAP and other observations for the central values
of $\mt$ and $\mb(\mb)$.
The total $\chi^2$ as a function of $m_{1/2}$ is shown in
\reffi{fig:CHI}. One can
see a global minimum of $\chi^2 \sim 4.5$
for both values of $\tb$. This is quite a good fit for the number of
experimental observables being fitted. 
Such a preference for not too heavy SUSY particles has also been found in
several other analyses~\cite{ehow3,ehow4,AllanachFit,mastercode}, see
also \citeres{Rosze,LSPlargeTB}. 

\begin{figure}[htb!]
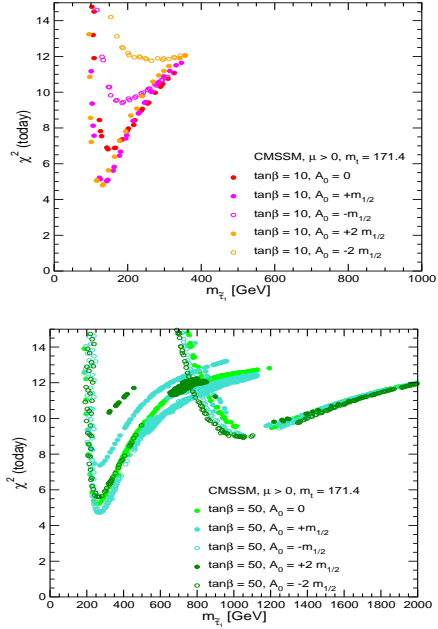

\begin{center}
\includegraphics[width=.4\textwidth,height=4cm]
                {ehow5.mass03a.1714.cl.eps}\\[3mm]
\includegraphics[width=.4\textwidth,height=4cm]
                {ehow5.mass03b.1714.cl.eps}
\vspace{-1em}
\caption{%
The mass of the lighter $\tilde \tau$ with its $\chi^2$ values~\cite{ehoww}.
}
\label{fig:mstau}
\end{center}
\vspace{-1em}
\end{figure}

The $m_{1/2}$--$\chi^2$ relation can be translated into a prediction of
SUSY masses. As an example \reffi{fig:mstau} shows the mass of the
lighter $\tilde \tau$ together with the corresponding $\chi^2$
value~\cite{ehoww}. For $\tb = 10 (50)$ the preferred value is 
$\tilde \tau_1 \approx 150 (250) \gev$. In this way the EWPO analysis
offers good prospects for the LHC and the ILC and possibly even for the
Tevatron. 
In a similar way also $\Mh$ with its corresponding $\chi^2$ can be
analyzed. The LEP limit of $114.4 \gev$ as a lower bound and the upper
bound of 
$\Mh^{\rm CMSSM} \lsim 127 \gev$~\cite{PomssmRep,asbs2} naturally
squeeze the $\Mh$ prediction into this interval. More interesting is
the case where the lower LEP bound is left out. In this case, using four
EWPO, four BPO and the CDM constraint a best-fit value for $\Mh$ of 
$\sim 110 \ldots 115 \gev$ (depending on $\tb$) was obtained~\cite{ehoww}.
This is substantially higher than the SM result of \refeq{MHSMfit}.

A fit as close as possible to the SM fit for $\MHSM$ (resulting in
\reffi{fig:blueband}) has been performed in \citere{mastercode}. 
All EWPO as in the SM~\cite{LEPEWWG} (except $\Ga_W$, which has a
minor impact) were included, supplemented by the CDM constraint in
\refeq{cdmexp}, the $(g-2)_\mu$ results in \refeq{delamu} and the
$\br(b \to s \ga)$ constraint.
The $\chi^2$ is minimized with respect to all CMSSM parameters for
each point of this scan. Therefore, $\De \chi^2=1$ represents the
68\% confidence level uncertainty on $\Mh$.
Since the direct Higgs boson search limit from LEP is not used in this
scan the lower bound on $\Mh$ arises as a consequence of 
{\em indirect} constraints only, as in the SM fit.

\begin{figure*}[htb!]
\begin{picture}(500,190) 
  \put(0,-10){ \resizebox{7.0cm}{!}
             {\includegraphics[angle=0]{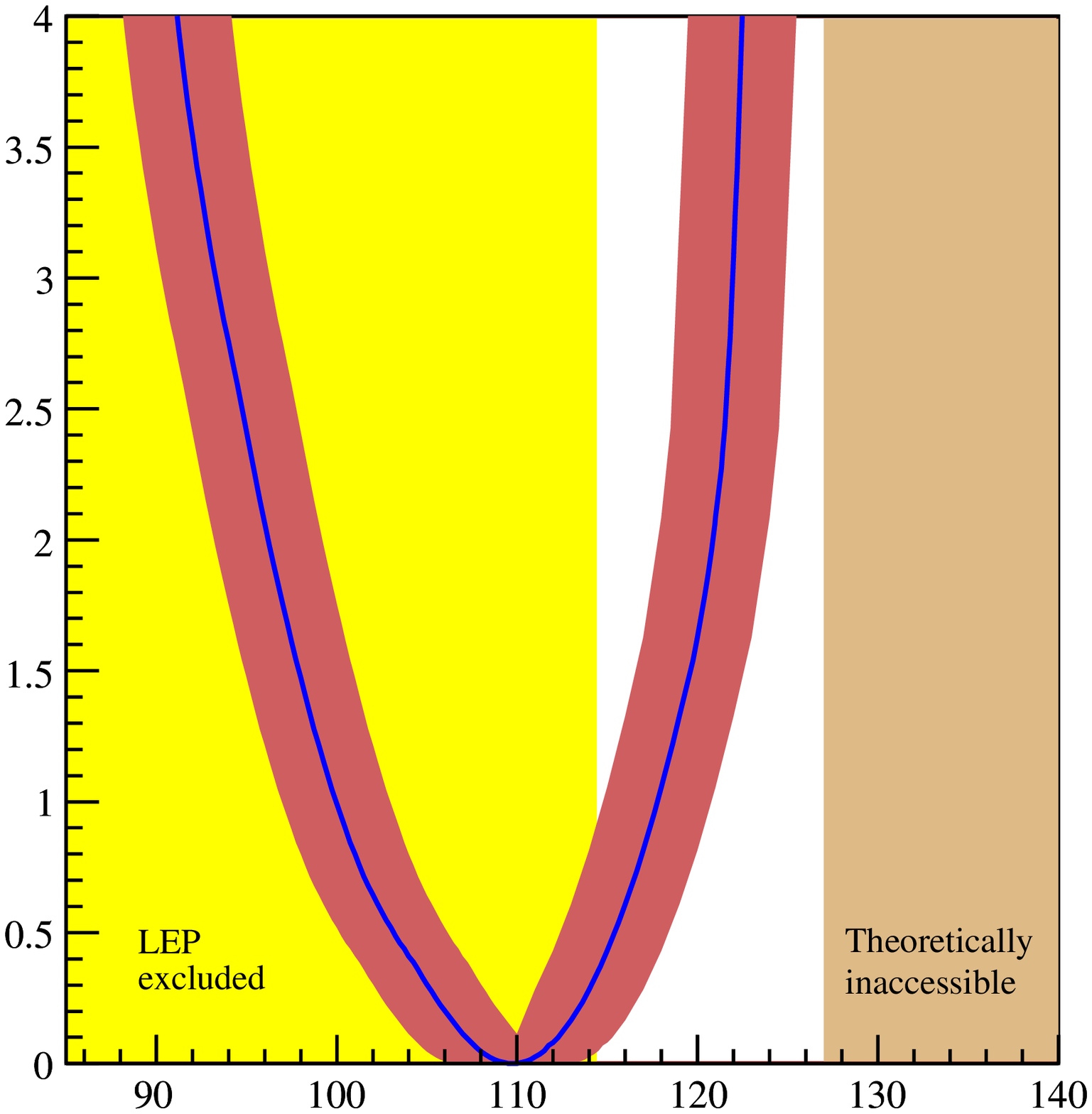}}  }
  \put(95, 200){CMSSM}
  \put(185, -10){$\Mh$ [GeV]}
  \put(00, 220){\begin{rotate}{90}$\Delta \chi^2$\end{rotate}}
  \put(310,-10){ \resizebox{7.0cm}{!}
               {\includegraphics{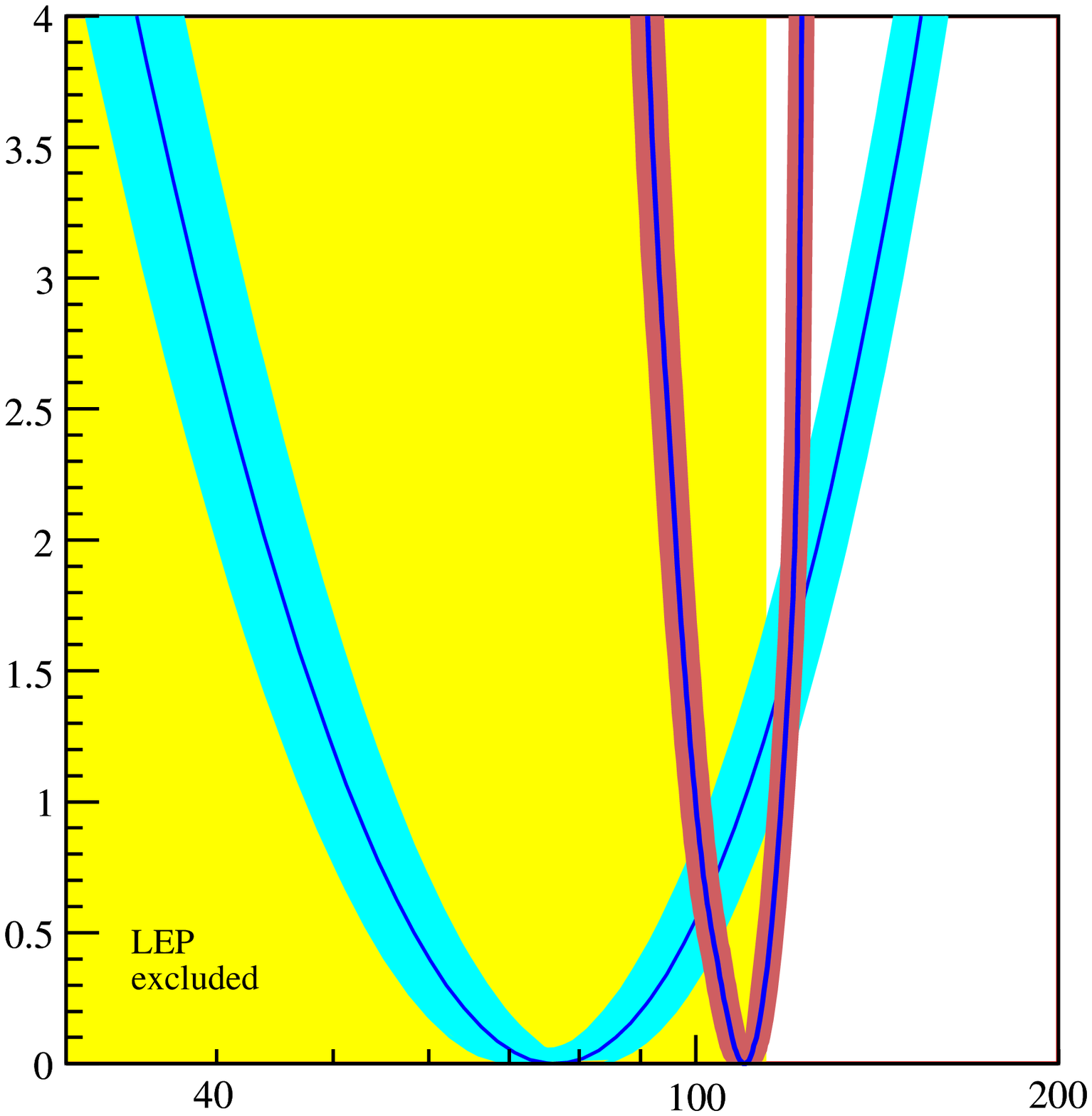}}  }
  \put(480, -10){$M_{\rm Higgs}$ [GeV]}
  \put(315, 220){\begin{rotate}{90}$\Delta \chi^2$\end{rotate}}
  \put(525, 40){CMSSM}
  \put(400, 200){SM}
\end{picture}
\caption{Left: Scan of the lightest Higgs boson mass versus $\De\chi^2$.
  The curve is the result of a CMSSM fit
  using all of the available constraints (see text). The direct limit
  on $\Mh$ from LEP~\cite{LEPHiggsSM,LEPHiggsMSSM} is not included.
  The red (dark gray) band represents the total
  theoretical uncertainty from unknown higher-order corrections,
  and the dark shaded area on the right above $127 \gev$ 
  is theoretically inaccessible (see text).
  Right:  Scan of the Higgs boson
  mass versus $\Delta \chi^2$ for the SM (blue/light gray), as
  determined by \cite{LEPEWWG} using all available electroweak
  constraints, and for comparison, with the CMSSM scan superimposed
  (red/dark gray). 
}
\label{fig:mh_vs_chi2}
\end{figure*}

In the left plot of \reffi{fig:mh_vs_chi2}~\cite{mastercode} the
$\De\chi^2$ is shown as a function of $\Mh$ in the CMSSM. The area
with $\Mh \ge 127$ is theoretically inaccessible, see above. The right
plot of \reffi{fig:mh_vs_chi2} shows the red band parabola from the
CMSSM in comparision with the blue band parabola from the SM. There is a
well defined minimum in the red band parabola, leading to a prediction
of~\cite{mastercode}  
\BE
\Mh^{\rm CMSSM} = 110^{+8}_{-10}\;{\small{\rm(exp)}} 
                     \pm 3\;{\small{\rm(th)}\gev ,}
\label{MH_CMSSM}
\EE
where the first, asymmetric uncertainties are experimental and the
second uncertainty is theoretical (from the unknown higher-order
corrections to $\Mh$~\cite{mhiggsAEC,PomssmRep}). 
The fact that the minimum in \reffi{fig:mh_vs_chi2} is
sharply defined is a general consequence of the MSSM, where the
neutral Higgs boson mass is not a free parameter as described above. 
The theoretical upper bound $\Mh \lsim 135 (127) \gev$ in the (C)MSSM
explains the sharper rise of the $\De\chi^2$ at large $\Mh$
values and the asymmetric uncertainty. In the SM, $\MHSM$ is a
free parameter and only enters (at leading order) logarithmically in
the prediction of the precision observables. In the (C)MSSM this
logarithmic dependence is still present, but in addition $\Mh$
depends on $\mt$ and the SUSY parameters, mainly from the scalar
top sector. The low-energy SUSY parameters in turn are all connected
via RGEs to the GUT scale parameters.  
The sensitivity on $\Mh$ in the analysis of \citere{mastercode} (and
also of \citere{ehoww}) is therefore the combination of the indirect
constraints on the four free CMSSM 
parameters and the fact that $\Mh$ is directly predicted in
terms of these parameters.   
%
%
This sensitivity also gives rise to the fact that the fit result in
the CMSSM is less affected by the uncertainties from unknown
higher-order corrections in the predictions of the electroweak
precision observables. While the theoretical uncertainty of the CMSSM
fit (red/dark gray band in \reffi{fig:mh_vs_chi2}) is dominated by
the higher-order uncertainties in the prediction for $\Mh$, the
theoretical uncertainty of the SM fit (blue/light gray band in
\reffi{fig:mh_vs_chi2}) is dominated by the higher-order
uncertainties in the prediction for the effective weak mixing angle,
$\sweff$~\cite{sw2eff2l}. 
The most striking feature is that even {\em without} the direct
experimental lower limit from LEP of $114.4 \gev$
the CMSSM prefers a Higgs boson mass which is
quite close to and compatible with this bound. From the curve in
\reffi{fig:mh_vs_chi2}, the value of the $\chi^{2}$ at the LEP
limit corresponds to a probability of 20\% (including theoretical
errors in the red band). This probability may be compared with the SM
with a 12\% $\chi^{2}$
probability at the LEP limit (including theoretical errors from the
blue band).




\subsection*{Acknowledgements}

\noindent
We thank the organizers of {\em Lepton Photon 07} for the invitation
and financial support.
We furthermore thank A.~Ritz, D.~St\"ockinger and G.~Weiglein for help
in the preparation of this talk and
M.~Gr\"unewald for helpful discussions.



\end{document}